# A Metamaterial-inspired Approach to RF Energy Harvesting

C. Fowler (cfowler@mail.usf.edu) — Department of Physics, University of South Florida, 4202 E. Fowler Avenue, Tampa, FL 33620

J. Zhou (jiangfengz@usf.edu)* — Department of Physics, University of South Florida, 4202 E. Fowler Avenue, Tampa, FL 33620

**Abstract**

We demonstrate an RF energy harvesting rectenna design based on a metamaterial perfect absorber (MPA). With the embedded Schottky diodes, the rectenna converts captured RF waves to DC power. The Fabry-Perot (FP) cavity resonance of the MPA greatly improves the amount of energy captured. Furthermore, the FP resonance exhibits a high Q-factor and significantly increases the voltage across the Schottky diodes, which improves the rectification efficiency, particularly at low power. This leads to factor of 16 improvement of RF-DC conversion efficiency at ambient intensity level.

**Introduction**

Low-power RF energy harvesting has become a field of interest due to the need to acquire power in situations where the use of wires and/or batteries is impractical such as in the cases of expansive sensor networks and structural health monitoring. The recent proliferation of RF signals due to cell phones, Wi-Fi networks, and GPS has produced readily available ambient power sources (although of low power density) for scavenging energy. Historically there has been difficulty with creating omnidirectional electrically small energy harvesting antennas at RF frequencies that efficiently collect energy[1]. There is further difficulty with efficient AC to DC rectification in cases of low input power[2]. In this paper we demonstrate the potential of metamaterial-based energy harvesting antennas to address both of these issues[3].

Research in metamaterials is a relatively new and unexplored field of study with objectives for constructing materials with customized optical, mechanical, or electrical parameters that would otherwise be difficult or impossible to obtain with ordinary materials[4,5]. Metamaterials are constructed by building a 2-D or 3-D array of structures within a substrate. The choice of the structures used in the array is the primary determinant of the properties of the bulk metamaterial. Thus the bulk properties of the metamaterial can be tuned by altering the shape, spacing, size, etc. of the structures that the metamaterial is composed of. Provided that the size of the structures is sufficiently smaller than the wavelength of the incident electromagnetic waves, the waves interact with the structures in a fashion analogous to the way that optical light or x-rays interact with the atoms and molecules of ordinary materials. There are a multitude of potential applications for metamaterials such as the construction of negative index materials[6,7], cloaking[8], perfect lenses[9], and energy harvesting[10] (which will be the focus of this paper).

A big limitation of metamaterials is that they will typically only function as intended within a limited bandwidth. The structures used to make a metamaterial typically resonate such that the metamaterial exhibits the desired properties only at its resonance frequencies. Thus the metamaterial functionality is often confined to a relatively small bandwidth surrounding each resonance peak, the performance of which depends on the Q-factor of the resonance.

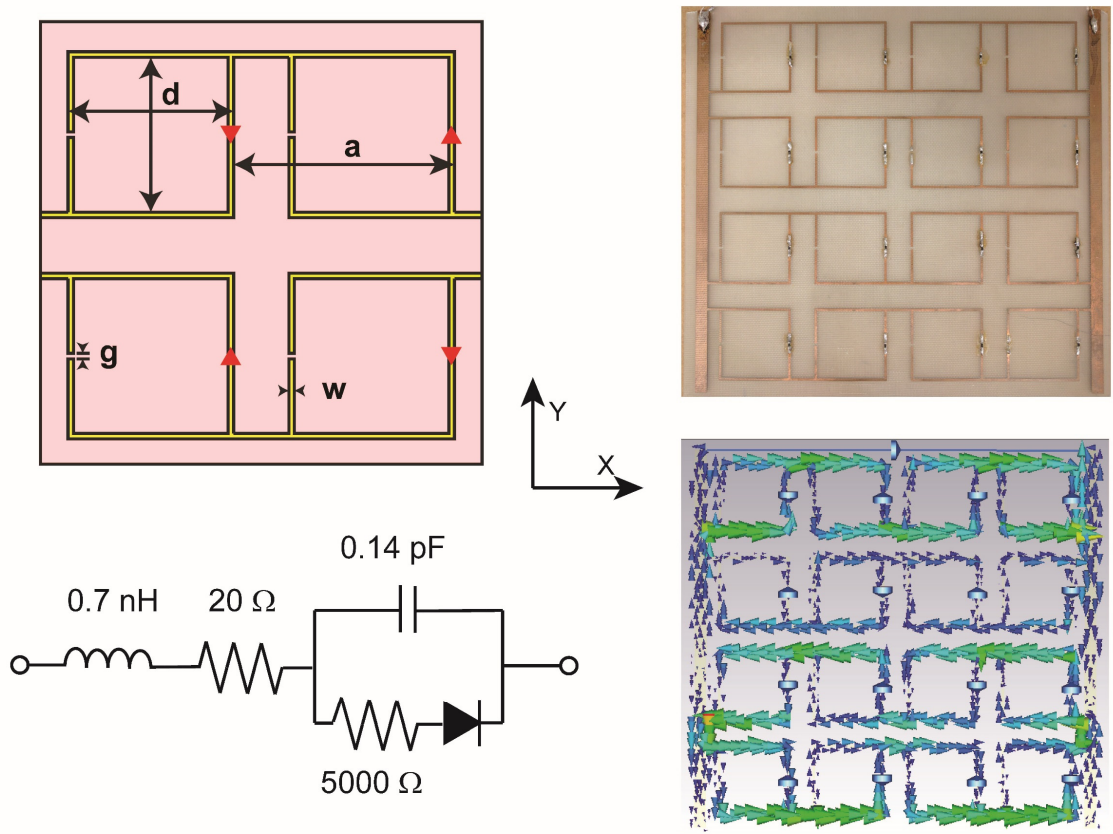

*Figure 1 - Rectenna sample design with the dimensions of the unit cell and the equivalent circuit model used for the Schottky diodes in simulations. Diodes are indicated in the unit cell by red triangles pointed in the direction of current flow. d = 30 mm, a = 20 mm, g = 1mm, w = 1mm, t = 0.8128 mm (thickness of FR4 substrate), s = 0.0178 mm (thickness of copper ring layer). The incident wave is perpendicular to the plane of the rectenna and polarized in the x-direction. The surface current modes activated at the resonance frequency (908 MHz) are shown on the sample.*

      An antenna constructed out of an appropriately designed metamaterial that incorporates a means of rectification can operate as an efficient energy harvesting system that can be used to convert power captured from RF sources (or possibly other regions from the electromagnetic spectrum) to useful DC power [11,12]. Optimal RF energy harvesting occurs when the metamaterial is constructed such that a high Q-factor resonance occurs at the same frequency as the external radiation source. For ambient RF energy harvesting, these frequencies will usually be found in the ISM band (particularly those from 800 MHz to 6 GHz) from sources such as cell phone signals, GPS, and WI-FI. References [13-15] provide measurements of power density levels that one might expect to find in urban environments. The greatest power density measured was 6.4 µW/cm$^2$, but average values typically ranged from 1-50 nW/cm$^2$ depending on the frequency, time of day, and proximity to RF sources.

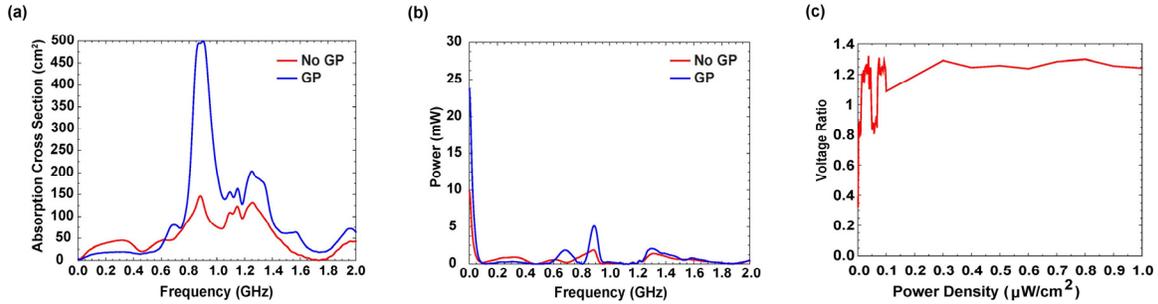

*Figure 2 - (a) Comparison of simulated rectenna absorption with and without the ground plane. The absorption peak narrows and increases, indicating an increase in the Q-value of the resonance. (b) Simulated power delivered to the load from a Gaussian pulse. The DC component of the power delivered to the load is the largest component, suggesting good rectification efficiency. The DC component is also strongly enhanced by the ground plane. (c) Oscilloscope measurement of the DC voltage across one of the Schottky diodes. The presence of the ground plane enhances the voltage, which makes it easier for the diode to activate (particularly at low power) and thus improves energy harvesting efficiency.*

Reference [16] gives a nice summary of past attempts at developing high efficiency low-power energy harvesting systems for various power densities and frequencies. As mentioned by the authors, the results between groups are not directly comparable due to differing frequencies, definitions of efficiency, and power densities. Nevertheless, they provide a reference point for evaluating the results presented by this paper. Other efficiency benchmarks are presented by references[17,18]. The results of our measurements on the energy harvesting efficiency of the metamaterial sample described in figure 1 are presented in this paper. Upon examination of the results presented by other groups, we believe these results to be the most efficient over the power densities and frequencies investigated.

**Results**

The absorption cross section obtained by simulation of the rectenna (figure 2a) predicts a peak energy harvesting frequency around 0.8 GHz. Adding the ground plane causes the peak to become somewhat narrower, but much larger (an increase in Q-value). As seen in figure 2b, the power delivered to the load is primarily from DC, while power from AC signals is much smaller. When the ground plane is incorporated, the DC power is much higher. A rough oscilloscope measurement of the voltage across one of the diodes shows the voltage-enhancing effect produced by inclusion of the ground plane (figure 2c).

Efficiency measurements for relatively high power densities (figure 3) indicate that the maximum energy harvesting efficiency occurs at 0.9 GHz, which is comparable to the frequency predicted by the absorption cross section. The power density range is well above what would typically be expected to be available from ambient RF signals and thus would only be found within close proximity (<100m) to a strong RF source such as a cell phone tower or right next to a weak RF source. In general, the efficiency of the sample increases with incident power density. However, when the ground plane is present, there is a point where the efficiency begins to diminish with increased power density (the red curve in figure 3f). When the ground plane is placed at the optimum distance (4cm), the energy harvesting efficiency improves considerably (compare figures 3a-3c with 3d-3f) and goes well over 100% at times. This indicates that the ground plane has caused the effective area of the rectenna to become larger than its geometric area.

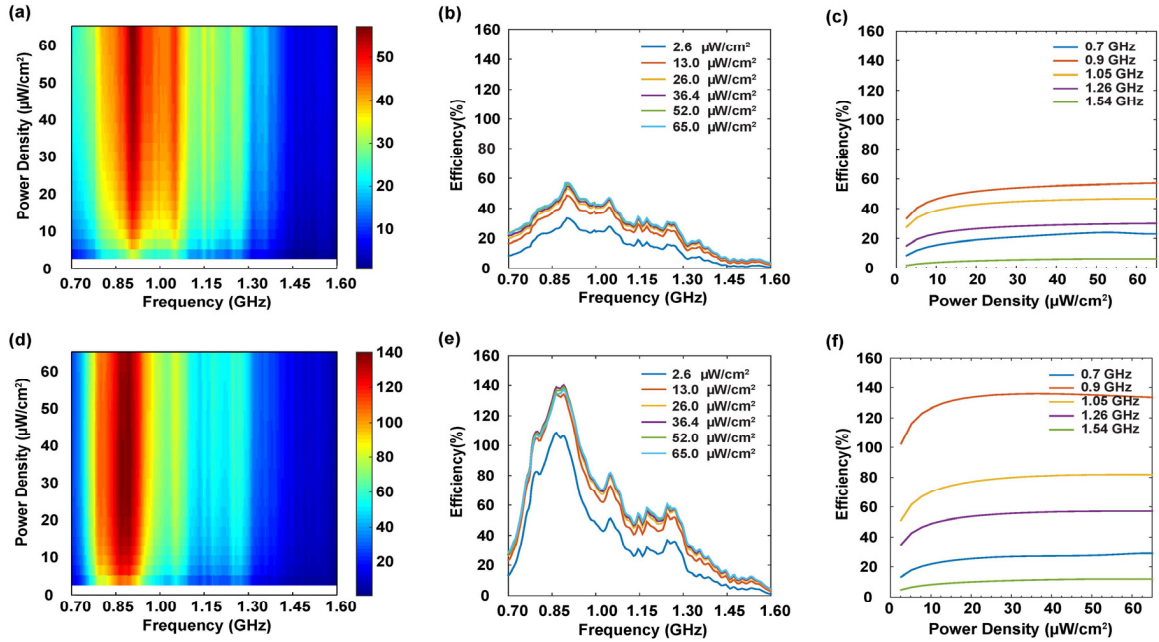

*Figure 3 - Experimental measurements of energy harvesting efficiency as a function of frequency and power density. (a)-(c) are measurements without the ground plane. (d) - (f) are measurements with the ground plane.*

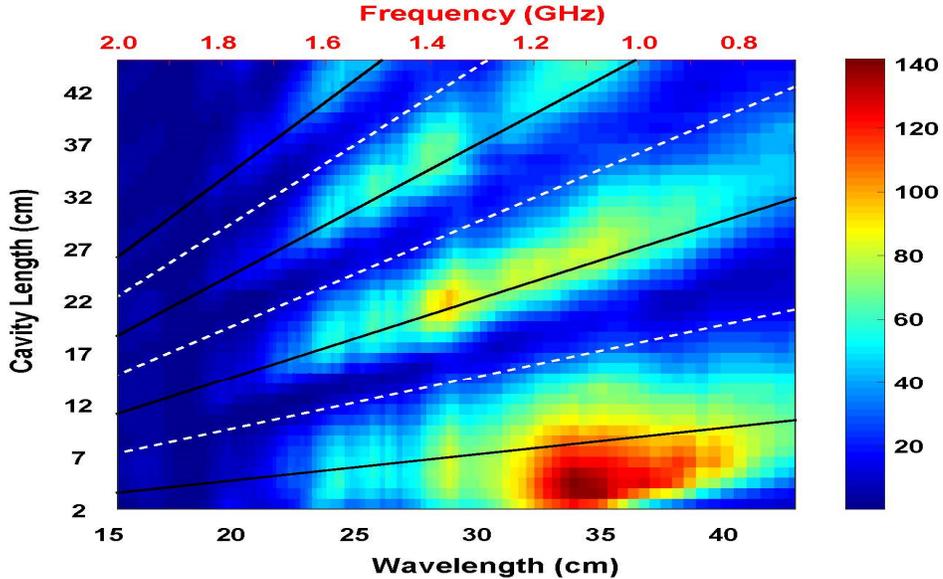

*Figure 4 - Measurement of the effect of Fabry-Perot cavity length on energy harvesting efficiency at fixed power density (30 µW/cm²). Solid Lines indicate the theoretical prediction where maximum energy harvesting efficiency would occur for a vacuum-filled cavity with half-open boundary conditions. Dashed lines indicate where energy harvesting maxima would occur for a vacuum-filled cavity with closed boundary conditions. The metamaterial perfect absorber rectenna behaves similarly to a half-open FP cavity. The shortest observed resonance length (λ/8) is less than predicted (λ/4), because the effective index of refraction of the metamaterial surface is larger than one.*

Figure 4 demonstrates the efficiency-enhancing effect of the Fabry-Perot cavity created by the rectenna and the ground plane. Examining the relationship between the efficiency, ground plane distance, and RF wavelength reveals standing wave patterns resembling that of a resonating cavity with one side closed (ground plane side) and the other side open (rectenna side). In the figure, the standing wave pattern is compared to the theoretical predictions of vacuum-filled Fabry-Perot cavities with half open (solid black lines) and closed (dashed white lines) boundary conditions.

Figures 5a and 5b show the energy harvesting efficiency measurements at power densities that are comparable to what can be found from ambient sources (up to 1.0 µW/cm$^2$). The power density levels presented were chosen to correspond to what can be found within relatively short distances from a commercially available high-power (100 mW) omnidirectional WIFI antenna with a maximum gain of 3/2. Now the efficiency is clearly much lower than it was for the high-power measurements. However, the presence of the ground plane increases the efficiency by a large factor (up to 16) in this case. A plot of the ratio of the efficiency with the ground plane to the efficiency without the ground plane, which will be referred to as the enhancement factor, is shown in figure 5c. The presence of the ground plane produces the largest enhancement factor at very low powers (about 0.01 µW/cm$^2$, which is at a typical ambient level), but unfortunately the overall efficiency at this level is still very low (only about 1%). At the lowest power density levels the efficiency is no longer enhanced much.

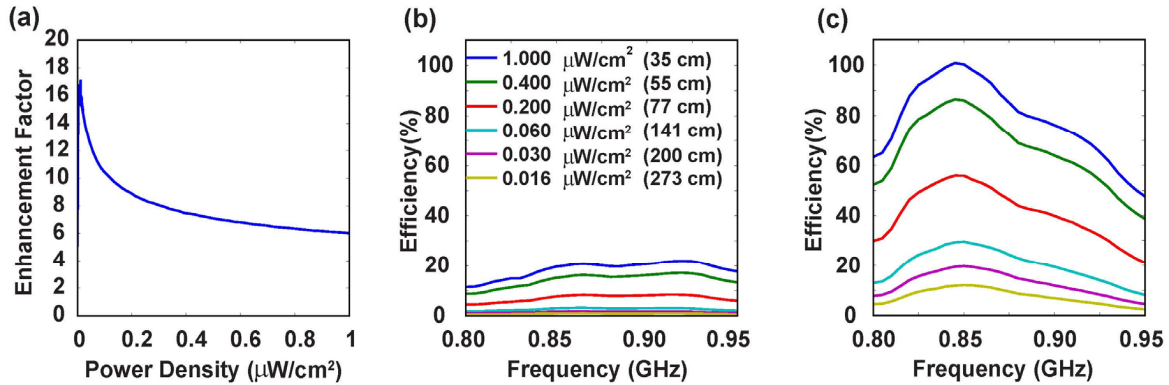

*Figure 5 - Measurements corresponding to power densities that would be available at the specified distances from a 100 mW source possessing a transmitting antenna with gain of 3/2. (a) No ground plane. (b) Ground plane present. Same legend as 5a. (c) Enhancement factor; defined as the ratio of power harvested with the ground plane to power harvested without the ground plane. Measured at 0.85 MHz. The peak frequency is slightly different than before because of slight differences in the angle of incidence of the rectenna between measurements.*

## Discussion

The data indicates several advantages to basing an energy harvesting antenna off of a metamaterial perfect absorber. First, building the diodes into the metamaterial surface results in highly-efficient rectification at the resonance frequency as determined in both the simulated results (figure 2) and the experimental results (figure 3). Alignment of the FP resonance with the natural resonance of the metamaterial surface leads to the best results. Second, the tunability of the metamaterial's dielectric constant allows the rectenna to be designed such that it is impedance-matched to free space, and hence has a reflection coefficient near zero. This improves the absorption and has the added advantage of allowing for a smaller rectenna. Since the surface behaves closely to an open-boundary for the FP cavity, the resonance length is shorter than for a closed boundary cavity ($\lambda/4$ compared to $\lambda/2$). The measured FP resonance peak occurred at a wavelength of $\lambda/8$ instead of $\lambda/4$ (figure 4), because the index of refraction of the material is not quite equal to 1. Third, as evidenced by the measurement in figure 2c, the conducting ground plane increases the voltage across each of the diodes in the rectenna. The effect gets stronger at low power (figure 5c) since it can provide enough of a voltage boost to activate the diodes when the diode threshold voltage would not ordinarily be attained. Once the power gets too low, then the diode threshold can not be met even with the ground plane present. If the power is too high (well above ambient levels), the diodes reach their breakdown voltage and efficiency decreases as shown in figure 3f. Fourth, the ground plane increases the directivity of the antenna (and thus also the effective area) for angles of incidence for which the metamaterial surface is located between the ground plane and the source. However, this comes at the cost of directionality since the ground plane will reflect away any signals coming from angles of incidence for which the ground plane is located between the rectenna and the source.

The point about the ground plane increasing the directivity is important, because it may not be feasible to build an omnidirectional energy harvesting device capable of collecting enough ambient RF power to operate practical devices. Energy harvesting efficiency is proportional to the effective area, which is a function of wavelength and directivity. Since the wavelength will be pre-determined by the source from which energy is being harvested, the only way to increase the effective area is by increasing the directivity. However, it is impossible to simultaneously increase the directivity in all directions because an increase in directivity for some directions necessitates a decrease in others. To demonstrate the difficulty of energy harvesting with a low-directivity antenna, consider a perfectly efficient short dipole antenna with a directivity of 1.5; a well-known omnidirectional antenna. The total amount of energy that can be harvested at 900 MHz with a relatively high incident power density of 0.1 $\mu W/cm^2$ is about 13 $\mu W$. This is well below the 100 $\mu W$ threshold that might be needed for typical devices as suggested by Visser[19]. In circumstances where the location of the ambient radiation source is known, it would be beneficial to instead use a highly directive receiving antenna aimed at the source to maximize the effective area.

Future improvements in the design will be aimed at increasing the effective area, and improving the rectification efficiency of the antenna. The effective area may be improved by considering alternate structures and geometrical arrangements. Alternate structures could allow for polarization-independent energy harvesting. It's possible that the SRRs in the sample may be placed too close together such that the capturing areas of adjacent rings overlap and thus reduce the amount of power captured by each ring. A higher amount of power captured per ring should make it easier for the diodes to activate. Array theory for antennas suggests that the SRR's should be placed about half a wavelength apart for optimal power capture. To improve rectification efficiency, it may be worthwhile to replace the Schottky diodes with spin diodes[20] due to their superior zero-bias resistance. Another possibility is to replace the diodes with more complicated rectifying circuits that will increase DC voltage such as the Greinacher circuit[17].

**Conclusion**

Designing RF harvesting rectennas based off of metamaterial perfect absorbers is a promising solution for collecting ambient RF energy in low power density environments because they are tunable, highly efficient, and electrically small. However, to operate practical devices, the efficiency needs to be improved further, particularly in regards to overcoming the zero-bias resistance to activate the diodes. The reflecting ground plane component of the metamaterial rectenna dramatically improves the RF-DC conversion efficiency, especially for lower available power densities, by helping to overcome the zero-bias resistance and by increasing the effective area of the antenna for certain angles of incidence. The biggest disadvantage of the ground plane is that it precludes omnidirectionality, although this may be unavoidable for effective energy harvesting.

**Method**

*Fabrication*

The metamaterial samples are fabricated using photolithography techniques. First, a mask that outlines a copper ring structure containing two gaps is designed and created. The dimensions of the unit cell and the design of the complete sample are shown in figure 1. The mask is placed on top of a copper-plated piece of printed circuit board composed of FR4 substrate which is then placed under an exposure lamp for 10 minutes. The mask is removed and the sample is washed with a diluted developer solution and then bathed in ferric chloride until any copper that was not shielded by the mask has been removed. After the excess copper has been removed, Skyworks SMS 7630 Schottky diodes are soldered across the right-hand gap of each ring with the orientations indicated by the red triangles on the unit cell shown in figure 1. The copper rings in the metamaterial capture RF waves, which are rectified by the diodes to deliver DC power to a load. The configuration of an antenna with a built-in rectification circuit is commonly referred to as a rectenna.

It is important to note that the rectenna sample is very sensitive to the polarization and angle of incidence of incoming RF signals. Since the sample is composed of an array of copper loops, electrical current can be driven by both the electric and magnetic fields. The orientation of the fields with respect to the sample determines which resonance modes will be excited and at which frequency they will occur. As such, deviations in the angle of incidence and/or polarization affect both the amount of energy harvested and the optimum frequency.

*Experimental Setup*

A signal generator produces a sinusoidal RF signal and sweeps both frequency and output power with frequencies ranging from 0.7-2.0 GHz. The signal is sent to an amplifier which increases the signal strength by about 45 dB (depending on the frequency). A small fraction of the amplified signal is extracted with a coupler, and then measured with a signal analyzer to determine the strength of the amplified signal. The amplified signal is transmitted by an SAS-570 horn antenna at normal incidence to the rectenna sample located about 380 cm away from the transmitting antenna. A metal ground plane is placed behind the sample with the purpose of creating a Fabry-Perot cavity to increase the amount of captured RF radiation. The experiment takes place within an anechoic chamber to isolate the rectenna sample from multi-path interference and outside signals. The rectenna sample is connected to a decade resistor box, which is used as a proxy for a device that DC power will be delivered to.

*Theory*

The DC power ($P_{DC}$) harvested from the sample can be determined simply by measuring the DC voltage (V) across the load with a multimeter and applying the well-known equation $P_{DC} = V^2/R$. A program written with LabVIEW is used to sweep over the appropriate frequency and power ranges while recording the voltage across the load to measure the efficiency (η) as described below. The efficiency is measured by taking the ratio of the rectenna's effective area to its geometric area. Effective area is defined as the ratio of the power absorbed at the load divided by the Poynting vector at the location of the rectenna and is given by[21]:

$$\frac{P_{DC}}{S} \qquad (1)$$

To determine the poynting vector at the location of the rectenna sample, the following equation from antenna theory is used[21]:

$$S = \frac{P_t \times G_t}{4\pi D^2} \qquad (2)$$

$P_t$ is the power radiated from the transmitting antenna, $G_t$ is the linear gain of the transmitting antenna, and D is the distance between the transmitting antenna and the sample. The gain of the transmitting antenna was determined by using the standard 3-antenna technique for gain measurements[21].

Putting this all together leads to the following definition which will be called the energy harvesting efficiency:

$$\eta = \frac{\frac{V^2}{R}}{S \times A_{geo}} \times 100\% \qquad (3)$$

This is actually the same way that aperture efficiency is defined for aperture antennas. For aperture antennas the efficiency will always be less than 100%, but for non-aperture antennas the effective area can be larger than the geometric area (a classic example of this is a dipole antenna). In such cases this definition of efficiency can yield results greater than 100%. It is important to recognize that this would not imply that the receiving antenna is capturing more power than is incident on it.

To characterize the efficiency of an energy harvesting system completely, the RF to DC conversion efficiency ($\eta_{RF-DC}$) should also be determined. $\eta_{RF-DC}$ indicates how much of the captured energy was successfully converted to DC power. This is the ratio of the DC power harvested to the RF power captured by the antenna.

$$\eta_{RF-DC} = \frac{P_{DC}}{P_{RF}} \times 100\% \qquad (4)$$

Due to the nonlinearity of the diodes contained within the rectenna, a way to determine $P_{RF}$ has not been determined, and so the focus of this paper is only on measuring the energy harvesting efficiency, η. Methods for determining $P_{RF}$ will be explored in future experiments.

*Experimental Concerns*

There are a few considerations concerning the accuracy of the data. First, standing waves are generated between the transmitting antenna and the sample. This will result in higher power transfer at some frequencies and lower power transfer at others for a given distance between the sample and the transmitting horn antenna. It's estimated that this affects the power received by the sample varying by as much as ±10%. However, this effect is minimal at the frequency for which the power transfer is most efficient, since this is when the sample's reflection coefficient is at its smallest. Hence, the measurements taken at the frequency of peak efficiency will be the most accurate. Second, there is some power transmitted to the sample due to a second harmonic generated by the amplifier used for the transmitting antenna. The harmonic transmits a maximum of 5% of the total power radiated from the transmitting antenna, which occurs at the largest power densities that can be produced by the system. However, the harmonic signals are of frequencies that are inefficiently absorbed by the sample and thus are a very small contribution to the absorbed power. This effect is essentially negligible even at the highest measured power densities. Lastly, the precision of the transmitting system is not as good at high power due to limitations of the power settings of the signal generator. This results in some efficiency measurements appearing artificially lower than they should be. This effect is negligible when the power density is under 35 $\mu W/cm^2$. It is also negligible for power densities up to 50 $\mu W/cm^2$ while the frequency is less than about 1.3 GHz.

Some simplifications to this experiment have been made that slightly reduced the accuracy of the results, but greatly eased taking measurements. Firstly, since the impedance of the diodes is a function of both frequency and power, the impedance of the load should be adjusted accordingly for each measurement to maximize the power transferred to the load. We found that when the load was set to 1000 ohms, that the deviations from optimal efficiency were very small with changes in frequency and power. Secondly, the inclusion of a power management circuit might be necessary for real world applications to compensate for fluctuations in the signal being harvested. This is left out of the experiment so that we can focus on the energy harvesting efficiency of the rectenna alone.

### *Simulations*

To supplement the experimental results, the absorption properties of the rectenna are simulated using CST Microwave Studio. Limitations in integrating circuit models into CST for plane wave simulations restricted us to using the Schottky diode model shown in figure 1, which is a simplified version of one used by Keyrouz et al[18]. The parameters for the model were obtained from the manufacturer's specifications. A broadband radar cross section monitor is used during a transient solver simulation of a plane wave to measure the absorption cross section (ACS) of the rectenna, which gives a rough indication of which frequencies the energy harvesting efficiency will be best. For the ACS calculation, the idealized diode in the equivalent circuit model is omitted to avoid generation of harmonics that interfere with the results. A surface current monitor is then used to identify the resonance modes of the rectenna at its peak energy harvesting frequency. Simulating the absorption cross section is primarily useful for estimating the peak frequencies for energy harvesting, but it is not good for predicting the correct proportionality for the efficiency across all frequencies. This is partly because of the limitations of the simple diode model used in the simulation. Another reason that the proportions may differ from measurements is that the absorption cross section includes energy losses in the rectenna, whereas the efficiency refers only to power absorbed in the load. As an additional test for energy harvesting capability of the sample, an incident Gaussian pulse was simulated and power delivered to a load was calculated. Although this technique will not indicate which frequency is best for producing DC power, it helps for getting an idea about how well the sample rectifies captured waves.

# References

1. Wheeler, H. A. Fundamental limitations of electrically small antennas. *Proceedings of the Institute of Radio Engineers* **35**, 1479-1484 (1947).
2. Cardoso, A. J., de Carli, L. G., Galup-Montoro, C. & Schneider, M. C. Analysis of the rectifier circuit valid down to its low-voltage limit. *IEEE Transactions on Circuits and Systems I-Regular Papers* **59**, 106-112 (2012).
3. Ziolkowski, R. W. & Erentok, A. Metamaterial-based efficient electrically small antennas. *IEEE Transactions on Antennas and Propagation* **54**, 2113-2130 (2006).
4. Shalaev, V. M. Optical negative-index metamaterials. *Nature Photonics* **1**, 41-48 (2007).
5. Soukoulis, C. M. & Wegener, M. Past achievements and future challenges in the development of three-dimensional photonic metamaterials. *Nature Photonics* **5**, 523-530 (2011).
6. Smith, D. R., Padilla, W. J., Vier, D. C., Nemat-Nasser, S. C. & Schultz, S. Composite medium with simultaneously negative permeability and permittivity. *Physical Review Letters* **84**, 4184-4187 (2000).
7. Veselago, V. G. Electrodynamics of substances with simultaneously negative values of epsilon and mu. *Soviet Physics Uspekhi-Ussr* **10**, 509-514 (1968).
8. Pendry, J. B., Schurig, D. & Smith, D. R. Controlling electromagnetic fields. *Science* **312**, 1780-1782 (2006).
9. Pendry, J. B. Negative refraction makes a perfect lens. *Physical Review Letters* **85**, 3966-3969 (2000).
10. Landy, N. I., Sajuyigbe, S., Mock, J. J., Smith, D. R. & Padilla, W. J. Perfect metamaterial absorber. *Physical Review Letters* **100** (2008).
11. Zhu, N., Ziolkowski, R. W. & Xin, H. A metamaterial-inspired, electrically small rectenna for high-efficiency, low power harvesting and scavenging at the global positioning system L1 frequency. *Applied Physics Letters* **99** (2011).
12. Ramahi, O. M., Almoneef, T. S., Alshareef, M. & Boybay, M. S. Metamaterial particles for electromagnetic energy harvesting. *Applied Physics Letters* **101** (2012).
13. Piñuela, M., Mitcheson, P. D. & Lucyszyn, S. Ambient RF energy harvesting in urban and semi-urban environments. *IEEE Transactions on Microwave Theory and Techniques* **61**, 2715-2726 (2013).
14. Visser, H. J., Reniers, A. C. F. & Theeuwes, J. A. C. Ambient RF energy scavenging: GSM and WLAN power density measurements. *EuMC 2008* **38th**, 721-724 (2008).
15. Bouchouicha, D., Dupont, F., Latrach, M. & Ventura, L. Ambient RF energy harvesting. *International Conference on Renewable Energy and Power Quality* (2010).
16. Falkenstein, E., Roberg, M. & Popovic, Z. Low-power wireless power delivery. *IEEE Transactions on Microwave Theory and Techniques* **60**, 2277-2286 (2012).
17. Hawkes, A. M., Katko, A. R. & Cummer, S. A. A microwave metamaterial with integrated power harvesting functionality. *Applied Physics Letters* **103** (2013).
18. Keyrouz, S., Perotto, G. & Visser, H. J. Frequency selective surface for radio frequency energy harvesting applications. *IET Microwaves, Antennas, and Propagation* **8**, 523-531 (2014).
19. Visser, H. J. & Vullers, R. J. M. RF energy harvesting and transport for wireless sensor network applications: principles and requirements. *Proceedings of the IEEE* **101**, 1410-1423 (2013).
20. Hemour, S. *et al.* Towards low-power high-efficiency RF and microwave energy harvesting. *IEEE Transactions on Microwave Theory and Techniques* **62**, 965-976 (2014).

21	Balanis, C. A. *Antenna Theory: Analysis and Design*. 3rd edn, pp. 89, 92, and 1031 (Wiley-Interscience, 2005).